\documentclass[%
 reprint,
superscriptaddress,
nofootinbib,
 amsmath,amssymb,
 aps,
]{revtex4-2}
\usepackage[]{graphicx}
\usepackage{dcolumn}
\usepackage{bm}
\usepackage[colorlinks=true, allcolors=blue]{hyperref}
\usepackage{booktabs}
\usepackage{subcaption}

\usepackage{braket}
\usepackage{amsmath, amsthm, amssymb}
\usepackage{comment}

\usepackage[dvipsnames]{xcolor}
\usepackage{ulem}

\usepackage{dsfont}
\usepackage{xcolor}
\usepackage{todonotes}
\usepackage{ulem}
\usepackage{listings}

\newcommand{\sampleFreq}{\mathrm{\Phi}_{\mathrm{sample}}}
\newcommand{\qubitFreq}{\mathrm{\Phi}_{\mathrm{qubit}}}
\begin{document}

\title{A time-to-digital converter with steady calibration through single-photon detection}

\author{Matías R.~Bolaños}
\affiliation{Dipartimento di Ingegneria dell'Informazione, Universit\`a degli Studi di Padova, Via Gradenigo 6B - 35131 Padova, Italy.}
    
\author{Daniele~Vogrig}
\affiliation{Dipartimento di Ingegneria dell'Informazione, Universit\`a degli Studi di Padova, Via Gradenigo 6B - 35131 Padova, Italy.}

\author{Paolo~Villoresi}
\affiliation{Dipartimento di Ingegneria dell'Informazione, Universit\`a degli Studi di Padova, Via Gradenigo 6B - 35131 Padova, Italy.}
\affiliation{Padua Quantum Technologies Research Center, Universit\`a degli Studi di Padova, via Gradenigo 6B, IT-35131 Padova, Italy}
\affiliation{Istituto Nazionale di Fisica Nucleare (INFN) -- sezione di Padova, Italy}

\author{Giuseppe~Vallone}
\affiliation{Dipartimento di Ingegneria dell'Informazione, Universit\`a degli Studi di Padova, Via Gradenigo 6B - 35131 Padova, Italy.}
\affiliation{Padua Quantum Technologies Research Center, Universit\`a degli Studi di Padova, via Gradenigo 6B, IT-35131 Padova, Italy}
\affiliation{Istituto Nazionale di Fisica Nucleare (INFN) -- sezione di Padova, Italy}

\author{Andrea~Stanco}
    \email{andrea.stanco@unipd.it}
    \affiliation{Dipartimento di Ingegneria dell'Informazione, Universit\`a degli Studi di Padova, Via Gradenigo 6B - 35131 Padova, Italy.}
    \affiliation{Padua Quantum Technologies Research Center, Universit\`a  degli Studi di Padova, via Gradenigo 6B, IT-35131 Padova, Italy}
    \affiliation{Istituto Nazionale di Fisica Nucleare (INFN) -- sezione di Padova, Italy}
    
\date{\today}

\begin{abstract}
Time-to-Digital Converters (TDCs) are a crucial tool in a wide array of fields, in particular for quantum communication, where time taggers performance can severely affect the quality of the entire application. 
Nowadays, FPGA-based TDCs present a viable alternative to ASIC ones, once the non-linear behavior due to the intrinsic nature of the device is properly mitigated.
To compensate for said nonlinearities, a calibration procedure is required, which should be maintained throughout its runtime.
Here we present the design and the demonstration of a TDC that is FPGA-based showing a residual FWHM jitter of 27 ps, that is scalable for multichannel operation. The target application in Quantum Key Distribution (QKD) is discussed with a calibration method based on the exploitation of single-photon detection that does not require stopping the data acquisition or using any estimation methods, thus increasing accuracy and removing data loss. 
The calibration was tested in a relevant environment, investigating the behavior of the device between 5 $^{\circ}$C and 80 $^{\circ}$C. Moreover, our design is capable of continuously streaming up to 12 Mevents/s for up to $\sim$1 week without the TDC overflowing making it ready for a real-life scenario deployment.
\end{abstract}

\maketitle

\section{Introduction}
Time-to-digital converters (TDCs) are used in a wide range of applications \cite{Henzler2010,tancock2019} where picosecond resolution is needed for time detection, such as high-energy particle physics \cite{Balla2014}, light detection and ranging (LiDAR) \cite{Zhang2019}, quantum communications \cite{Shen2013,Avesani2021,Carrier2023} and quantum random number generation \cite{Stanco2020}. 
Recently, Field-Programmable Gate Arrays (FPGAs) have emerged as a popular platform for the implementation of TDCs due to their flexibility, high-speed performance, and cost-effectiveness \cite{wang2010fully,Markovic2013,SANO201750,Garzetti2021}. 
However, general FPGA-based TDCs require periodic calibration to maintain accuracy \cite{Ito2010,Chen2019,Wang2022}. TDCs can be calibrated with a process known as the \textit{code density} test, which estimates the propagation time of each individual bin by sampling a uniformly distributed signal over time\cite{prasad2022,Rivoir2016Fully,Rivoir2006Statistical}. 
This is typically done using an internal ring oscillator and requires pausing data acquisition, which limits its use in continuous or time-sensitive applications.
To mitigate this, some systems estimate delay-line drift over time based on changes in the ring oscillator frequency after an initial calibration \cite{Ito2010,Wang2022}, though this introduces inaccuracies under non-uniform thermal drift. 
Other solutions include dual-chain calibration architectures \cite{jachna2014permanently}, which enable continuous calibration but double the logic resources needed per channel. More recently, event-driven calibration approaches have emerged, where real measurement events are used to update the calibration dynamically \cite{Won2016}. These methods offer real-time adaptation without acquisition downtime and form the basis for the steady calibration strategy adopted in this work.

One relevant application where time-to-digital conversion capabilities highly affect performance of the whole application is Quantum Key Distribution. Quantum Key Distribution (QKD) allows two parties (Alice and Bob) to create a shared, secret key. 
To do so, Alice encodes the information in degrees of freedom of a given quantum system, usually photons \cite{Scarani2008}. 
To make the sharing of the key successful, Bob requires precise timing of arrival of photons to synchronize both parties effectively and to allow better distinguishability between signal and noise. 
For example, algorithms like Qubit4Sync allow Bob to reconstruct Alice's clock signal from his own qubit measurements \cite{Calderaro2020}. 
Moreover, QKD is currently evolving to be implemented using satellite links to increase the distance achievable with these protocols, which is also a requirement for a global quantum network \cite{Sidhu2021, Pirandola2019}. 
This implies that communication could only be done while the satellite passes through the ground station's field of view. 
During this limited time, Bob should receive as much data as possible while maintaining its performance, implying that both calibrating or losing calibration mid-measurement can considerably hinder key generation.

To address the limitations of existing calibration techniques, we propose an FPGA-based TDC, named MARTY and implemented on a Zynq7020 chip, that integrates a calibration method we refer to as \textit{steady calibration}. This method is part of the broader class of event-driven calibration techniques, where the system self-corrects based on the same detection events it is designed to timestamp. While similar ideas have been explored in positron emission tomography (PET), where calibration updates are triggered by gamma detections from radioactive sources \cite{Won2016}, we adapt the concept to the context of quantum communication. Specifically, our system uses the natural stream of single-photon events from a pulsed laser and single-photon detector, such as those already present in a QKD setup, to continuously refresh its calibration without interrupting acquisition.

The paper has the following structure: in the following Subsection we provide a comparison over different calibration methods, in Section \ref{sec:materials} we present the working principle of the system, including the FPGA design, normal and steady calibration procedures, and finally how we performed temperature variation tests on the device; in Section \ref{sec:results} we present the results obtained in this work, including a QKD test performed with MARTY; eventually, in Section \ref{sec:conclusion} we present the final conclusions and future perspectives.

\subsection{Calibration method comparison}

Several calibration strategies have been developed for FPGA-based TDCs, each with distinct trade-offs. The most widely used method is the static code-density test, where a uniformly distributed signal, commonly from a ring oscillator (RO), is used to populate a histogram of bin counts \cite{prasad2022,Rivoir2016Fully,Rivoir2006Statistical}. From this, relative bin widths are inferred. While accurate under controlled conditions, this method requires stopping the acquisition, making it unsuitable for continuous or time-critical systems.

An alternative is drift estimation, which begins with a static calibration and later applies corrections by tracking changes in the frequency of the reference oscillator \cite{Ito2010,Wang2022}. This method avoids data loss but assumes homogeneous environmental effects across all bins, which may lead to cumulative inaccuracies during runtime, especially under thermal gradients.

Dual-chain calibration architectures achieve real-time correction by dedicating a second delay line for continuous monitoring of environmental drift \cite{jachna2014permanently}. This ensures accurate calibration without interrupting data acquisition, but at the cost of a significant increase in logic and routing resources, often doubling the requirements per channel, which is an impractical solution for systems with many channels or limited FPGA fabric.

The class of event-driven calibration methods offers a promising alternative by using measurement data itself to update calibration parameters \cite{Won2016, Whitehead2023, Jin2025}. This feature not only enhances the efficiency of the system, but also improves the overall time accuracy as each measurement is also a contribution to calibration and there is no distinction between ``data acquisition" and ``calibration" phases. In \cite{Won2016}, for example, every detection event from a PET gamma source is used to recalibrate the TDC in real time. In a very recent work by Jin et al. \cite{Jin2025}, the authors provide a thorough review of the different continuous calibration methods, focusing on a continuous calibration scheme that exploits signals coming from an entanglement source. However, in this work we show the possibility to use single photon events, allowing for a broader applicability. In fact, our steady calibration is an application of the same principles, tailored for QKD and photon-counting experiments, where high timing precision is of the utmost importance \cite{Shen2013, Terhaar2023, Castelvero2024}. Instead of relying on radioactive decay, or entangled photons, we use weak coherent pulses detected by a single-photon detector, which are already part of the experimental protocol. This makes the method hardware-efficient and well-suited for long-duration, low-resource, or environmentally variable applications, such as satellite quantum key distribution. In particular, this last application would be greatly benefited from this calibration method, since the measurement times are constrained by the satellite passage, and the harsh environment within orbit greatly decreases accuracy without adding a continuous calibration scheme.

\section{Materials and methods}\label{sec:materials}
\begin{figure*}[t!]
    \centering
    \includegraphics[width=2.0\columnwidth]{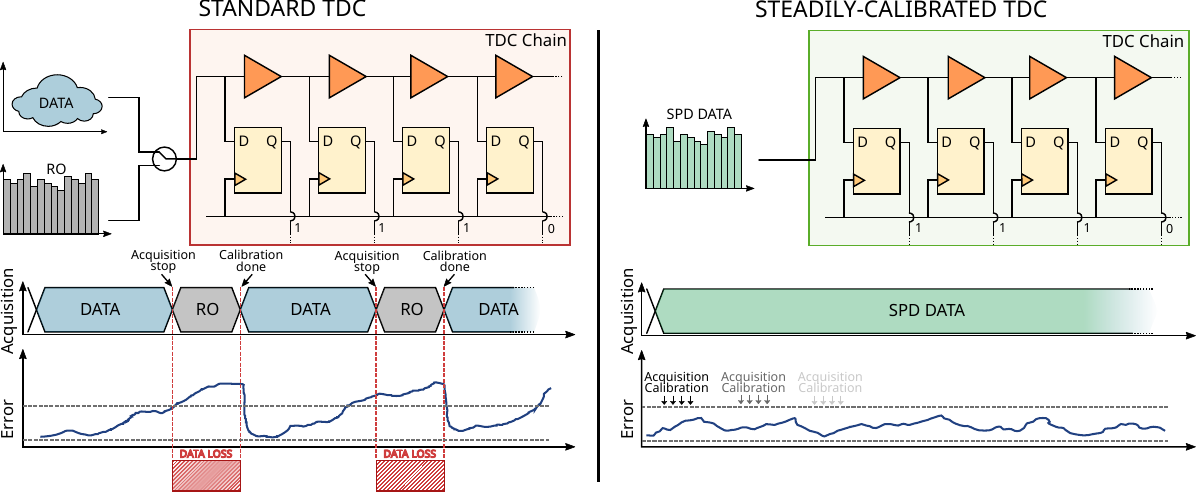}
    \caption{Qualitative functioning principle of a steadily-calibrated TDC (right) compared with a normal TDC (left).}
    \label{fig:main-scheme}
\end{figure*}
\subsection{Design of the device}

A generic TDC channel can consist of two components: a clock counting module for the coarse time estimation and a tapped delay line (TDL) for the fine estimation (Fig. \ref{fig:main-scheme} left).

While the clock counting module is a binary counter that increases at each rising edge of a clock, the TDL is a ``propagation line" where the electric signal travels through a chain of similar elements, each one with an intrinsic propagation delay. The implementation of such structure can be realized in different ways. In MARTY, for the coarse counting module, the DSP48 components were used as they grant high efficiency and speed, clocked by a $f_{\rm s}=412.5$~MHz clock, $\sampleFreq$. 
With the given clock frequency and the choice of encoding the coarse counter in 48 bits, MARTY has a total runtime of $2^{48}/f_{\rm s}\sim1$ week before overflow. 

The TDL was implemented through a chain of fast-carry components, the CARRY4 element (each containing 4 fast-carry). This fast-carry component was chosen due to its availability within the Zynq7000 chip family. To move the design to a different chip family, like Ultrascale+, a new fast-carry component has to be found and chosen.
In particular, for the Ultrascale+ family, a viable fast-carry component is the CARRY8, analogous to the Zynq7000's CARRY4 \cite{Castelvero2024}.
As usual in TDC implementation, each point of the TDL is flip-flopped to capture the current state at each clock edge. This snapshot output can be considered as a thermometer code\footnote{A thermometer code is a code of length $N_c$ consisting of the first $n_1$ elements equal to 1, followed by $n_0=N_c-n_1$ elements equal to 0.} that characterizes the propagation time inside the TDL and can be translated into a binary code using a dedicated decoder. 
The decoding algorithm used in MARTY was the one presented by \textit{Adami\v{c} et al.} in \cite{Adamic2019}, which, in turn, is based on a pipelined adder tree \cite{Wang2017}. This algorithm was chosen due to its high robustness against `bubble errors' without increasing the complexity of the system. 
`Bubbles' behave like a `bit swap' between two neighboring bits of the thermometer code (e.g. a code that should be 11111000 instead appears as 11110100), and they appear due to a variety of effects such as metastability on the flip-flops used to snapshot the TDL state \cite{Henzler2010}. 

It is worth noticing that the propagation time inside each fast-carry component is not uniform, ranging from units of picoseconds to around $100$~ps. 
Therefore, a calibration is needed to precisely characterize such propagation times.
By using 36 CARRY4 elements (144 fast-carry components in total), the coarse period $\tau_{\rm coarse}=1/f_s\simeq 2.42$~ns was fully covered. Indeed, a maximum length of $N_c\in[129, 135]$ bins was obtained in all data acquisitions. For this characterization, we consider the beginning of the carry-chain as the first bin with non-zero propagation time.

The final timestamp is obtained by combining the binary value of the TDL thermometer code, named \textit{fine value}, and the value obtained from the clock counting module, named \textit{coarse value}. The output timestamps of a channel can then be stored in a block RAM (BRAM) module to later transfer them to an external PC. 

\begin{figure}
    \centering
    \includegraphics[width=\columnwidth]{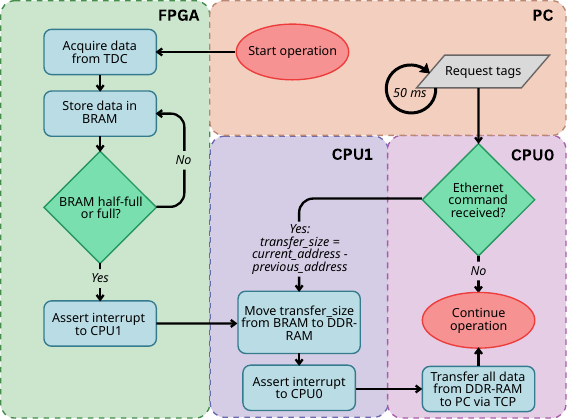}
    \caption{Flowchart of data transfer process from FPGA to PC to achieve 12 Mevents/s without bottlenecks.}
    \label{fig:data-flow-marty}
\end{figure}

The streaming architecture of the device, based on the one presented in \cite{Stanco2022}, takes advantage of the System-on-a-Chip (SoC) technology, exploiting not only the FPGA but also the CPU(s) counterpart to implement a continuous data transfer from the BRAM to the PC via the Ethernet protocol. 
With this scheme, the data is stored in the BRAM in a \textit{double-buffer}, where the CPU is interrupted whenever the BRAM reaches its half and full size, triggering the CPU to extract the time-tag information from the corresponding half. 
The size of the BRAM was carefully decided such that the time taken by the CPU to extract half the memory via Central Direct Memory Access (CDMA) lower than the time taken to fill the other half of the BRAM. Thus, the CPU is not interrupted to transfer a new block while the previous block is still transferring. 
This scheme allows to continuously stream the TDC data from the FPGA to the PC, with the most relevant limitation being the speed of the Ethernet cable of 1 Gbps, bringing to a maximum acquisition speed of $16.7$ Mevents/s which gets brought down to $\sim$12 Mevents/s when taking into account data transfer overhead\footnote{To reduce overhead, bare-metal applications were preferred over OS-based ones.}. 
Furthermore, to facilitate software integration, a request-based scheme was also implemented, as it is a standard approach on commercial TDCs to transfer data to a PC. 
In this scheme, the acquired data is sent only when receiving an external request from a PC. 
Moreover, if a request is received, the system checks the current address in the memory and extracts just the areas of the memory between the previous requested address (which in case of interrupts from the BRAM are half and full size) to the current address. 
Following the same logic as before, a request time of 50 ms gives enough time for the CPU to extract at most half the size of the BRAM without slowing the system down.
Thus, it is possible to reach nearly the maximum data transfer rate which is close to 12 Mevent/s. A scheme of the data flow for this system is presented in Figure \ref{fig:data-flow-marty}.

\begin{figure*}[t!]
    \centering
    \includegraphics[width=2\columnwidth]{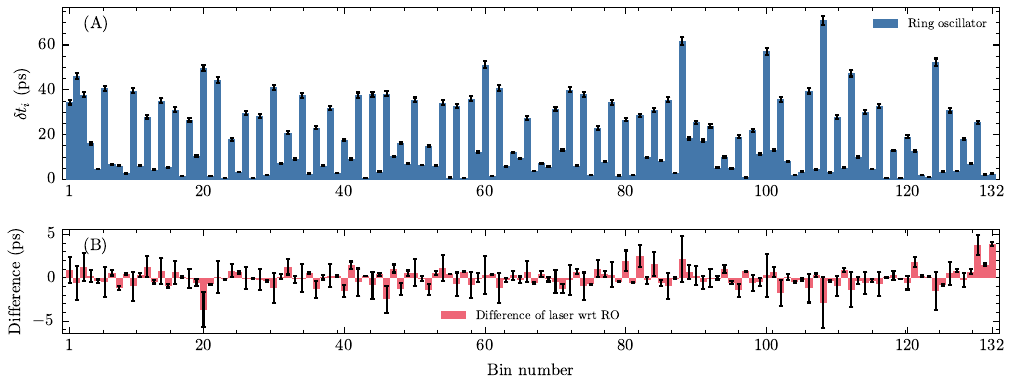}
    \caption{(a) Histogram obtained with a code density test using an FPGA-based RO as input. (b) Difference of the histogram obtained with a code density test using a laser detection signal with respect to the RO one.}
    \label{fig:code-density-comparison}
\end{figure*}

\subsection{System calibration and auto-calibration}\label{sec:calibration}

The calibration method is based on the evaluation of the results of a \textit{code density} test. 
A code density test consists of using a signal uniformly distributed over the time spanned by the TDL ($1/f_s\approx 2.42$~ns), and measuring the number of occurrences obtained from each bin. If the input signal is uniformly distributed and enough events are recorded, the occurrences in each bin should be proportional to its propagation time \cite{tancock2019}. 
As a standard solution, the code density test is performed using  an independent clock signal, such as an internal Ring Oscillator (RO)\footnote{A ring oscillator consists of an unstable loop of delay elements (usually not-gate) that generates a periodic signal not related to the system's clock, which can be considered uniformly distributed in time.}, that generates a uniformly-distributed signal connected to the input of the TDC. 
To obtain the minimum number of events needed to calibrate, $N_{\rm events}$, a method based on the one proposed in \cite{doernberg1984full} was used, such that
\begin{equation}
    N_{\rm events}>\left(\frac{z_{\alpha/2}}{\beta}\right)^2\cdot N_c,
\end{equation}
where $z_{\alpha/2}$ corresponds to the area under the standard normal distribution curve of $\alpha/2$, $\beta=0.1$ represents a tolerance of 10\% and $\alpha=0.02$ represents a confidence interval of $98$\%,  from which the required number of events $N_{\rm events}>75744$ was obtained. To simplify subsequent bitwise operations,  the amount of accumulated events was set to the closest power of 2 so that $N_{\rm events} = 131072 > 75744$.

From the number of events detected in the bin $i$, $w_i$, it is possible to derive the propagation time of the bin $\delta t_i=\frac{w_i}{\sum_j w_j}\tau_{\rm coarse}$, with $i\in[1,N_c]$ (Fig. \ref{fig:code-density-comparison}) and the corresponding cumulated propagation time $t_i$ as
\begin{equation}
t_i=
     \begin{cases}
     0 & \quad\text{if $i=0$}
     \\
     \sum_{j=1}^{i}\delta t_j
       &\quad\text{if }i\geq 1, \\
     \end{cases}
\end{equation}
which corresponds to assigning the edge of each bin as its cumulated propagation time value.
As it is common practice for calibration of TDC systems since it minimizes RMS errors \cite{khaddour2023calibration}, we assigned the calibrated time of bin $i$, $t^{(\rm c)}_i$, to the center of the bin instead of the edge, such that 
\begin{equation}
\label{eq:t_calibrated}
    t^{(\rm c)}_i = 
     \begin{cases}
     0 & \quad\text{if $i=0$}
     \\
       \frac12(\delta t_1+\delta t_{N_c}) &\quad\text{if $i=1$},\\
       \frac12\delta t_i+\sum_{j=1}^{i-1}\delta t_j
       &\quad\text{if }i>1, \\
     \end{cases}
\end{equation}
\noindent where in both cases we defined $t_0=t^{(c)}_0=0$ so that the difference between consecutive bins is well-defined for the entire carry-chain.
Definition \eqref{eq:t_calibrated} corresponds to assigning the center of each bin as its calibrated time value. This definition is mostly standard for \textit{code density} based calibration schemes for TDCs, but for this work we define the first calibrated bin in a different manner, including an extra term related to the last bin of the TDC chain:
the propagation time of the first bin is affected by the remaining propagation time of the last bin from the previous clock cycle. 
This re-definition was made such that the average difference between the time associated to consecutive bins is maintained constant and equal to the system's resolution $\tau_{\mathrm{res}}=\frac{\tau_{\rm coarse}}{N_{c}}$, independent whether the cumulative propagation time $t_i$ or the calibrated time $t_i^{(\rm c)}$ is used\footnote{Please note that, despite $t_1^{(\rm c)}$ does not exactly represent the center of the bin, we observed $\delta t_{N_c} \ll \tau_{\rm res}$, which means that $t_1^{(c)}\approx\frac{\delta t_1}{2}$.}. Indeed, by defining the difference between consecutive calibrated times $\delta t^{(c)}_{i} = t^{(c)}_{i} - t^{(c)}_{i-1}$, it holds that
\begin{equation}
\begin{aligned}
    \langle \delta t^{(c)}\rangle&=\frac{1}{N_c}\sum_{i=1}^{N_c} \delta t^{(c)}_i
    =\frac{1}{N_c}\left(\sum_{i=1}^{N_c}t^{(c)}_{i} - t^{(c)}_{i-1}\right)=\\
    &=\frac{1}{N_c}\left[\frac12\left(\delta t_1+\delta t_{N_c}\right)+\sum_{i=1}^{N_c}\frac12(\delta t_i + \delta t_{i-1})\right]\\
    &=\frac{1}{N_c}\sum_{i=1}^{N_c}\delta t_i = \langle \delta t\rangle=\frac{\tau_{\rm coarse}}{N_c}=\tau_{\rm res}. 
\end{aligned}
\end{equation}
\noindent This type of calibration requires the user to stop the current acquisition for about 30 ms, switching from the input signal to the RO one. 

In this work, we apply an extension of this method based on \cite{Won2016}, called \textit{steady calibration}, which allows the system to continuously calibrate while the device is acquiring data (Fig. \ref{fig:main-scheme} right). 
This method consists of storing the code density histogram in an ordered manner (i.e. the first element corresponds to the one detected first), so that at each new detected event, the first element is deleted and the new one is appended at the end.
The calibration curve is then recalculated with the new data. 
With this method, the system gets re-calibrated in a detection-by-detection basis, allowing it to detect variations on the order of up to tens of ns. 
At system startup, the system performs a standard static code density calibration to be the initial calibration value, which gets re-calculated and updated after every detection event.
With the chosen value for $N_{\rm events}$, assuming a commercial QKD system with detections in the order of $\sim400$ kevent/s, even though the calibration table gets updated in an event-by-event basis, a complete refresh of the table takes approximately $ 330$ ms.

The steady calibration method allows the exploitation of the acquired TDC data to calibrate itself in real time, assuming that the detected events are uniformly distributed over the TDL. 
Considering a pulsed laser as an event source with the repetition rate $f_q$ disciplined by a clock $\qubitFreq$, the temporal distribution is expected not to be uniform in the TDL outputs only if the two clocks $\qubitFreq$ and $\sampleFreq$ are commensurable, or locked to each other, and perfectly stable without any drift or jitter. 
Therefore, the combination of non commensurable clock frequencies and drift effects can assure a uniform distribution of the input events.

To test the proposed method, we obtained the calibration curve using a laser source and a single-photon detector (SPD) in place of the integrated RO. When comparing both histograms obtained with a code density test (Fig. \ref{fig:code-density-comparison}), the difference between both is considered negligible, with a $\chi^2=0.026$. From this, we can conclude that using an external single-photon detector with a laser signal is equivalent to the integrated RO in the FPGA chip for calibration purposes.
For the purpose of these tests, the calibration method was elaborated \textit{offline}.

\subsection{Temperature test setup}\label{sec:temperature-test}
The validity of the steady calibration method was verified by forcing a change in the working point of the system, which should require a calibration. Therefore, the device was tested at different temperatures as this can have a relevant impact on the performance of a TDC and is considered strictly related to the TDC calibration process \cite{Carra2018}. By modifying the temperature of the system in a controlled and fast manner, the reliability of the calibration method against changes in the environment conditions was tested. This is equivalent to testing the system over extended periods of time on a slower-changing environment.
To modify the device temperature, we used a combination of a Peltier cell and an RAL 9006 climatic chamber by Angelatoni Test Technologies. The combination of both temperature control systems allowed the FPGA chip to reach a wider temperature range, from 5~$^{\circ}$C to 80~$^{\circ}$C \footnote{The scanned range could not be further increased due to the climatic chamber not including humidity control. Thus, to protect the board integrity, we remained over 0~$^{\circ}$C}. To measure the temperature of the chip, the included XADC sensor of the board was used, which gives a temperature measurement in the center of the chip itself. 

The FWHM jitter and resolution of a two-channel implementation of MARTY was used to characterize its performance. To calculate the FWHM jitter, an input signal is divided into the two channels, which we assume equal for this purpose, and the histogram of the differences between the timetags is obtained, from which the FWHM jitter can be obtained. For this approximation, we assume that both channels have the same FWHM jitter and that the jitter of the source does not affect the resulting histogram due to both channels receiving the same input signal.

To perform the temperature measurements, a 1550 nm attenuated laser source was used as an input to a single-photon detector (PMD-IR detector by MPD), which was then divided and used as input for both channels of the TDC. The chip temperature was scanned in the whole range of 5~$^{\circ}$C to 80~$^{\circ}$C with a step of 1~$^{\circ}$C, where for each temperature the FWHM jitter and resolution were evaluated.

\subsection{Quantum Key Distribution test setup}

To evaluate the device performance for QKD applications, while also comparing it with commercial TDCs, a QKD transmitter using a three-state one-decoy BB84 protocol at $R=50$ MHz repetition rate, based on the source presented in \cite{Berra2023}, was used to encode an ``HVDD" sequence\footnote{An ``HVDD" sequence is a sequence of four qubits encoded using single-photon with Horizontal-Vertical-Diagonal-Diagonal polarizations.}. We used a standard polarization receiver scheme with four PMD-IR detectors by MPD, where for the time-tagger we used both a QuTAG from QuTools and MARTY. To evaluate the performance, a detection peak was measured with both devices, and the Quantum Bit Error Rate (QBER) for the HVDD sequence was estimated.

\section{Results and discussions}\label{sec:results}

\begin{figure}
    \centering
    \includegraphics[width=\columnwidth]{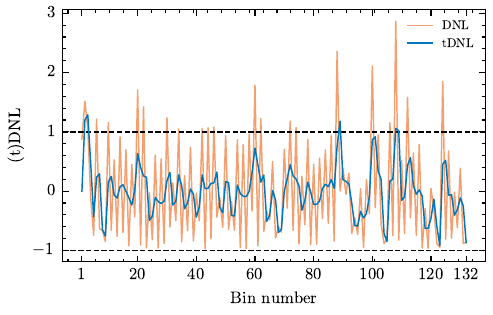}
    \caption{$\text{DNL}$ (in orange) and $\text{tDNL}$ (in blue) obtained for a single channel of the TDC at room temperature.}
    \label{fig:DNL-45C}
\end{figure}

\begin{figure}
    \centering
    \includegraphics[width=\columnwidth]{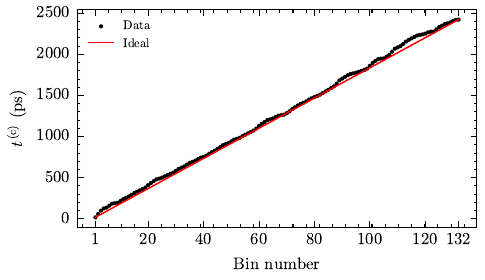}
    \caption{Calibrated time $t^{(c)}$ as a function of the bin number of the TDC, in orange the ideal behaviour for a uniform delay line.}
    \label{fig:resolution}
\end{figure}

MARTY's main performance metrics, such as resolution, differential nonlinearity (DNL), and FWHM jitter, were characterized using a two-channel TDC configuration \cite{Kalisz2004}.
As first instance, the mean resolution of the system was evaluated at room temperature such that
\begin{equation}
\label{eq:tau_res}
    \tau_{\mathrm{res}}=\frac{\tau_{\rm coarse}}{N_{c}}=18.37~\text{ps}.
\end{equation}
\noindent The DNL represents the deviation of the delay line with respect to an ideal one (uniformly distributed in time) \cite{ieee_standard}.
The DNL can be evaluated through a code density test with an internal RO as the input of the TDC, representing a signal uniformly distributed over the time at the input of the TDC. From the obtained counts for bin $i$, $w_i$ (Fig. \ref{fig:code-density-comparison}a), the DNL can be estimated as
\begin{equation}\label{eq:DNL}
    \text{DNL}_i=\frac{w_i-\langle w\rangle}{\langle w\rangle}=\frac{\delta t_i-\langle\delta t\rangle}{\langle\delta t\rangle},
\end{equation}
where $\langle w\rangle=\frac{1}{N_c}\sum_{i=1}^{N_c} w_i$, $\langle\delta t\rangle=\frac1{N_c}\sum_i \delta t_i=\tau_{\rm res}$ and $N_c$ is the last bin for which counts were recorded (Fig. \ref{fig:DNL-45C}).
A $\text{DNL}=[-0.97,\,2.86]$ was obtained, but at room temperature only 21 out of the 132 bins had $\text{DNL}>1$. 
Moreover, after obtaining the calibrated times $t^{(c)}$ (Fig. \ref{fig:resolution}), a non-linearity quantity, named $\rm tDNL$, is defined as 

\begin{equation}
    {\rm tDNL}_i =\frac{\delta t^{(c)}_{i}-\langle \delta t^{(c)}\rangle}{\langle \delta t^{(c)}\rangle}
\end{equation} 

where $\delta t^{(c)}_{i}$ corresponds to the difference between consecutive calibrated time bins and $\langle \delta t^{(c)}\rangle=\tau_{\rm res}$ as defined in Section \ref{sec:calibration}.
It is worth noting that ${\rm tDNL}_i$ does not provide additional information regarding the linearity of the system\footnote{Please note that this quantity is not a DNL by the standard definition.}, but instead provides a linearity notion `as seen by the user', i.e. a quantity that is relevant for the final user of the device, by utilizing the center of the bins instead of the edges of the bins for the DNL calculation.

\begin{figure}
    \centering
    \includegraphics[width=\columnwidth]{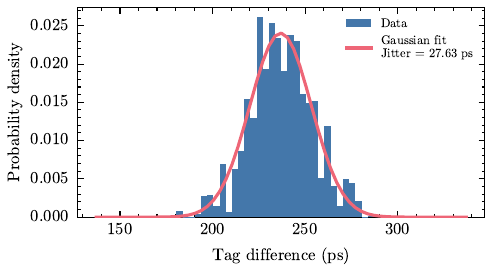}
    \caption{Histogram of the difference between two consecutive timetags in a two-channel implementation. In red, the corresponding Gaussian fit, from which the FWHM jitter was estimated.}
    \label{fig:jitter} 
\end{figure}

Lastly, to characterize the FWHM jitter of the system, a scheme like the one described in Section \ref{sec:temperature-test} was used, paired with a square wave as an input for both channels at room temperature. 
To obtain the FWHM jitter, a Gaussian fit was applied to the histogram of the difference between consecutive tags, from which the FWHM jitter was estimated as the $\text{FWHM}=27.63$~ps (Fig. \ref{fig:jitter}). 

\begin{figure}[!t]
    \centering
    \includegraphics[width=\columnwidth]{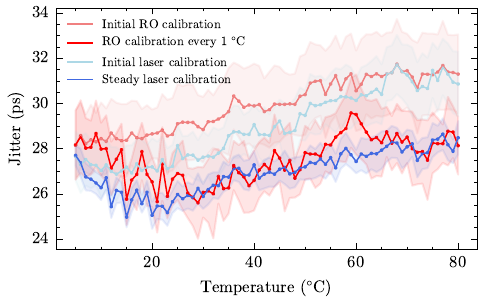}
    \caption{Comparison of the FWHM jitter of the system as a function of temperature obtained for different calibration methods: with the RO at 5~$^{\circ}$C and maintained through the scan (light orange); with the RO every 1~$^{\circ}$C step (red); with the single photon detector at 5~$^{\circ}$C and maintained through the scan (light blue); and steadily with the single photon detector (blue).}
    \label{fig:jitter-vs-temperature} 
\end{figure}

To determine the performance of the device with respect to changes in the environment, a temperature characterization of the system was performed using the setup described in Section \ref{sec:temperature-test}. 
The data acquisition process consisted on storing, for each temperature, a file with the calibration curve for each channel along with the registered tags in noncalibrated format with their respective channel. 
This format allowed to analyze the data with or without calibration, while also letting the possibility to implement calibration at particular points in the measurement. 
The acquired data for each temperature was processed in four different ways:
\begin{enumerate}
    \item Calibrated using the calibration curve obtained with the RO at 5~$^{\circ}$C (light orange curve in Fig. \ref{fig:jitter-vs-temperature}).
    \item Calibrated using the first time tags registered with the single photon detector at 5~$^{\circ}$C (light blue curve in Fig. \ref{fig:jitter-vs-temperature}).
    \item Calibrated using the calibration curve obtained for each temperature (red curve in Fig. \ref{fig:jitter-vs-temperature}).
    \item Steadily calibrated (blue curve in Fig. \ref{fig:jitter-vs-temperature}). 
\end{enumerate}

\begin{figure}
    \centering
    \includegraphics[width=\columnwidth]{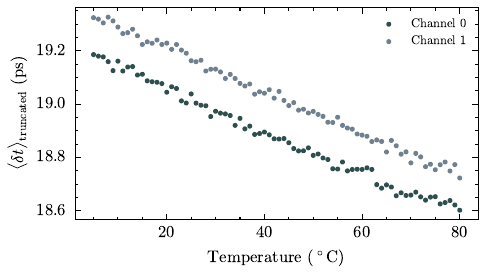}
    \caption{Mean single-bin propagation time for each channel of the TDC as a function of the temperature of the chip. The different offset between both channels was attributed to non-uniformity of components in the chip.}
    \label{fig:resolution-vs-temperature}
\end{figure}
As a first step, the relation between temperature and the average propagation time within the delay line, $\langle \delta t\rangle$, was analyzed.
Since the maximum number of registered bins is temperature and channel dependent (obtaining around 129 bins for 5~$^{\circ}$C and up to 135 for 80~$^{\circ}$C), we restricted this analysis to the first 120 elements of the delay line, such that
\begin{equation}
\langle \delta t\rangle_{\rm truncated}=\frac{\sum_{j=1}^{120}\delta t_j}{120}.
\end{equation}

The observed behavior depends on the particular architecture and materials of the Zynq7020 chip (28 nm process), of which not much information is publicly available, concluding that the CARRY4 element has a propagation time that is, on average, inversely proportional to temperature (see Fig. \ref{fig:resolution-vs-temperature}). 

Then, the FWHM jitter was calculated in the 4 scenarios previously mentioned (Fig. \ref{fig:jitter-vs-temperature}). 
For both cases in which calibration was done just at 5~$^{\circ}$C, increasing the chip temperature monotonically increased the FWHM jitter of the system (as also verified by \textit{Alshehry et al.} in \cite{Alshehry2023}), which highlights the importance of the steady calibration scheme, since transistors' characteristics are temperature dependent. 
On the other hand, when applying a calibration for every temperature step, the FWHM jitter decreased until 30~$^{\circ}$C, which implies an improvement in performance, and then increased until it stabilized at 60~$^{\circ}$C, close to the starting value. 
Finally, the one obtained with the steady calibration had a very similar shape to the one obtained by calibrating at each temperature, but with two notable improvements. In first instance, the system's FWHM jitter was found to be more stable, presenting an average standard deviation of $\langle\sigma_{\text{steady}}\rangle=0.61$~ ps, compared to the $\langle\sigma_{\text{RO}}\rangle=1.33$~ps obtained for the case when RO calibration was performed before each measurement. Moreover, as previously mentioned, the steady calibration does not require the acquisition to be stopped at any time, thus allowing the system to not lose any data.

Finally, a test with a QKD polarization-encoded source and receiver scheme was performed. 
The width of the detected pulse depends on the width of the optical pulse itself, the jitter of the detector, and the jitter of the time tagger. 
In real-world applications of QKD protocols, it is a common practice to apply time-filtering over the detected quantum signals such that the background noise is reduced. 
Therefore, the  jitter plays a relevant in order to lower the background counts, that are typically uniformly distributed in time. 
Indeed, the receiver jitter (that combines optical pulse width, single photon detector jitter and TDC jitter) determines the 
width of the temporal filter at the receiver.
A large receiver jitter corresponds to a 
large time-selection window, increasing the noise and the QBER.
For both devices, the mean QBER obtained was of $2.2$~\%\footnote{The QKD source and receivers were not properly optimized for performance as it is outside the scope of this work.}, showing that our design presents good performance for QKD applications. 
Even though the device was tested for a specific protocol (three-state one-decoy BB84), it is also applicable to all those QKD protocols that rely on time-tagging single photon events, such as discrete-variable including measurement device independent and twin-field protocols. 
The same can be said for the steady calibration, since all compatible protocols include a single-photon or weak coherent pulse source, which satisfy the signal requirement to be used for calibration purposes.

\begin{figure}[th!]
    \centering    
    \includegraphics[width=\columnwidth]{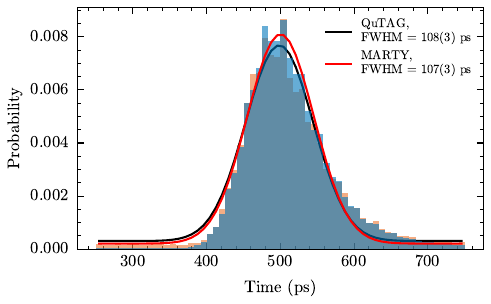}
    \caption{Pulse shape obtained with MARTY (orange bars) and the QuTAG (blue bars), with their corresponding Gaussian fit.}
    \label{fig:pulse-comparison}
\end{figure}

We compared the width of the detected pulse using both MARTY and the QuTAG, a high-level commercial time tagging system with 13.4 ps FWHM jitter in the standard version (Fig. \ref{fig:pulse-comparison}). 
It was observed that, even though Marty's FWHM jitter was higher than the QuTAG, the pulse widths obtained with both devices were comparable. This behavior is due to most of the width of the pulse coming from the detector's FWHM jitter, which is considerably higher than both time-taggers ($\sim100$ ps). 
Moving this design to a higher-level FPGA technology (e.g., Ultrascale+ with 16 nm FinFET process) can considerably increase the device's performances, which would not be beneficial when using InGaAs detectors such as the ones used in this work, but would make a noticeable difference when using superconductive nanowire single photon detectors, where the FWHM jitter is on the order of tens of picoseconds \cite{Castelvero2024, Wang2016}.

\section{Conclusion}\label{sec:conclusion}

An FPGA-based time-to-digital converter, named MARTY, was implemented on a Zynq7020 FPGA, coupled with a calibration method that increases detection accuracy and removes data loss. 
This calibration method exploits single-photon detection, allowing it to calibrate the device at each new event and without stopping data acquisition. 
MARTY is capable of acquiring data continuously for up to one week of runtime before overflowing, with parameters compatible with state-of-the-art results, those being resolution $\tau_{\mathrm{res}}=18.37$~ps, $\text{DNL}=[-0.97,\,2.86]$, and FWHM jitter $\tau_{\mathrm{jitter}}=27.63$~ps. 
The steady calibration method was tested between $5~^{\circ}$C and $80~^{\circ}$C, highlighting its validity as the FWHM jitter is kept under control while the temperature changing. 
Indeed, the steady calibration could be exploited in those environment that affect TDC performances becoming a key point of the dedicated application, like Satellite Quantum Communication.
Moreover, MARTY was tested and compared with a high-level commercial TDC using a QKD setup, where both the laser pulse FWHM and the QBER obtained with both devices were equal. 
For future developments, an \textit{online} version of the steady calibration feature could be implemented, while also upgrading to an Ultrascale+ (16 nm FinFET) based technology. This last point will increase the resolution and FWHM jitter even further thanks to a smaller process.
\bibliography{biblio}

\section*{Acknowledgment}
    M.R.B.W. acknowledges support from the European Union’s Horizon Europe Framework Programme under the Marie Sklodowska Curie Grant No. 101072637, Project Quantum-Safe Internet (QSI).

    M.R.B.W. acknowledges funding from the italian government via a study scholarship granted by the Ministry of Foreign Affairs and International Cooperation (MAECI) A.Y. 2021-2022.

    This work is (partially) supported by ICSC – Centro Nazionale di Ricerca in High Performance Computing, Big Data and Quantum Computing, funded by European Union – NextGenerationEU; Agenzia Spaziale Italiana, project Q-SecGroundSpace (Accordo n. 2018- 14-HH.0, CUP: E16J16001490001); Agenzia Spaziale Italiana (2020-19- HH.0 CUP Grant No. F92F20000000005, Italian Quantum CyberSecurity I-QKD).

    The authors would like to thank Dr. S. Bonora for the usage of the climatic chamber and Mr. A. Vanzo for the help in setting it up.

\end{document}